%% file: K_matrix_emulator.tex
\documentclass[10pt,aps,prc,floatfix,twocolumn,nofootinbib,superscriptaddress]{revtex4-1}
\usepackage{graphicx,amsmath,amssymb,bm}
\usepackage{amsfonts}
\usepackage[utf8]{inputenc}
\usepackage{verbatim}
\usepackage{float}
\usepackage[labelfont={small},font=small,subrefformat=parens,caption=false]{subfig}
\usepackage{cancel}
\usepackage{multirow}
\usepackage{array}
\usepackage{xparse}
\usepackage{xspace}

\usepackage{bigstrut}

\usepackage{physics}
\usepackage{color}
\usepackage{dsfont}

\usepackage{dcolumn}

\usepackage[pdfencoding=auto, pdfpagelabels]{hyperref}

\usepackage[overload]{textcase}

\definecolor{linkcolor}{rgb}{0,0,0.40} 
\hypersetup{%
    pdfsubject=Paper,
    pdfkeywords={nuclear physics} {Bayesian} {chiral EFT} {emulator},
    unicode = true,
    breaklinks = true,
    colorlinks = true,
    linkcolor = linkcolor,
    citecolor = linkcolor,
    menucolor = linkcolor,
    urlcolor = linkcolor
}

\usepackage{cellspace}
\graphicspath{{./figures/}}

\input{buqeye_macros}  

\begin{document}

\title{Fast \& accurate emulation of two-body scattering observables without wave functions}

\author{J.~A. Melendez}
\email{melendez.27@osu.edu}
\affiliation{Department of Physics, The Ohio State University, Columbus, OH 43210, USA}

\author{C. Drischler}
\email{drischler@frib.msu.edu}
\affiliation{Facility for Rare Isotope Beams, Michigan State University, MI 48824, USA}

\author{A.~J. Garcia}
\email{garcia.823@osu.edu}
\affiliation{Department of Physics, The Ohio State University, Columbus, OH
43210, USA}

\author{R.~J. Furnstahl}
\email{furnstahl.1@osu.edu}
\affiliation{Department of Physics, The Ohio State University, Columbus, OH 43210, USA}

\author{Xilin Zhang}
\email{zhang.10038@osu.edu}
\affiliation{Department of Physics, The Ohio State University, Columbus, OH 43210, USA}

\newcommand{\T}{K}
\newcommand{\Tritz}{\widetilde \T}
\newcommand{\Tfunc}{\mathcal{\T}}
\newcommand{\Tfuncopt}{\mathcal{\T}_{\star}}
\newcommand{\NEC}{\ensuremath{n_{t}}}
\newcommand{\lec}{a}
\newcommand{\coeff}{\beta}
\newcommand{\coeffs}{\vec{\coeff}}
\newcommand{\coeffsopt}{\vec{\beta}_{\star}}

\newcommand{\regwf}{J}
\newcommand{\irrwf}{Y}

\newcommand{\regfreewf}{j}
\newcommand{\irrfreewf}{y}

\newcommand{\verifyvalue}[1]{#1}
\newcommand{\eg}{\textit{e.g.}\xspace}
\newcommand{\ie}{\textit{i.e.}\xspace}

\date{\today}

\begin{abstract} 

We combine Newton's variational method with ideas from eigenvector continuation to construct a fast \& accurate emulator for two-body scattering observables. The emulator will facilitate the application of rigorous statistical methods for interactions that depend smoothly on a set of free parameters. Our approach begins with a trial $K$ or $T$ matrix constructed from a small number of exact solutions to the Lippmann--Schwinger equation. Subsequent emulation only requires operations on small matrices. We provide several applications to short-range potentials with and without the Coulomb interaction and partial-wave coupling. It is shown that the emulator can accurately extrapolate far from the support of the training data. When used to emulate the neutron-proton cross section with a modern chiral interaction as a function of 26 free parameters, it reproduces the exact calculation with negligible error and provides an over 300x improvement in CPU time.  

\end{abstract}

\maketitle

\section{Introduction} \label{sec:intro}

Nuclear scattering experiments yield invaluable data for testing, validating, and improving theoretical models such as chiral effective field theory (EFT)~\cite{Epelbaum:2008ga,Machleidt:2011zz,Hammer:2019poc,Epelbaum:2019kcf}---the method of choice for deriving microscopic nuclear interactions at low energies. 
However, there are competing formulations of chiral EFT 
with open questions on issues including EFT power counting, sensitivity to regulator artifacts, and differing predictions for medium-mass atomic nuclei.
Low-energy nuclear scattering data combined with rigorous statistical methods such as Bayesian parameter estimation~\cite{Wesolowski:2018lzj}, model comparison~\cite{Phillips:2020dmw}, and sensitivity analysis~\cite{Ekstrom:2019lss} applied to chiral EFT predictions will provide important insights to address these issues~\cite{Furnstahl:2021rfk}.

But taking full advantage of the available data using such statistical methods requires fast \& accurate predictions across a wide range of model parameters.
While, in principle, the scattering equations can be solved accurately in few-body systems, doing so is prohibitively slow for statistical analyses of three- and higher-body scattering, and even for two-body scattering more efficient alternatives are appealing.

In this Letter, we study such an alternative for two-body scattering that has the potential of future extensions to higher-body systems. We introduce an efficient emulator of the Lippmann--Schwinger (LS) integral equation using Newton's variational method~\cite{newton2002scattering,Rabitz:1973nm} combined with ideas from eigenvector continuation (EC)~\cite{Frame:2017fah,Sarkar:2020mad}.
The term emulator refers here to an algorithm capable of approximating the exact solution of a scattering problem with high accuracy while requiring only a fraction of the computational resources.

The power of EC as an emulator stems from the fact that, as the Hamiltonian parameters are varied, the trajectory of each eigenvector remains within a small subspace compared to the full Hilbert space.
Linear combinations of eigenvectors spanning this subspace are extremely effective trial wave functions for variational calculations (see also the reduced basis method~\cite{RHEINBOLDT1993849,chenRBA2017}).
Emulators based on EC have accurately approximated ground-state properties such binding energies and charge radii, and even transition matrix elements~\cite{Konig:2019adq, Ekstrom:2019lss, Wesolowski:2021cni,Yoshida:2021jbl}.
Additionally, EC has recently been used to construct effective trial wave functions for applying the Kohn variational principle to emulate two-body scattering observables~\cite{Furnstahl:2020abp}, and for $R$~matrix theory calculations of fusion observables~\cite{Bai:2021xok}.
As we will show in this Letter, Newton's variational method has the feature that scattering observables can be predicted using trial scattering matrices (\eg, the $\T$~matrix) rather than trial wave functions.
But emulated wave functions can still be obtained~\cite{newton2002scattering,taylor2006scattering}.

The remainder of this work is organized as follows.
In Sec.~\ref{sec:formalism} we briefly describe the formalism underlying the emulator.
We then present in Sec.~\ref{sec:results} several applications to short-range potentials with and without  the long-range Coulomb interaction and partial-wave coupling.
In addition to phase shifts, we study the neutron-proton ($np$) total cross section by combining multiple emulators across a set of (coupled) partial waves to assess the accuracy and speedup of the emulator in realistic scattering scenarios.
We conclude this Letter in Sec.~\ref{sec:summary}, and refer to the
\verifyvalue{Appendices} for more technical details including the emulation of gradients required by some Monte Carlo samplers and optimizers.
We use natural units in which $\hbar = c = 1$. 
The self-contained set of data and codes that generates all results shown in this Letter
\verifyvalue{will be made} publicly available~\cite{BUQEYEgithub}.

\section{Formalism} \label{sec:formalism}

We aim to construct an efficient emulator for the LS equation given a short-range potential $V(\lecs)$ that depends smoothly on a set of parameters $\lecs$, such as the low-energy couplings of a chiral potential.
Specifically, we consider here the LS equation for the scattering $\T$~matrix, which reads in operator form\footnote{
All subsequent equations work for any boundary conditions imposed via $G_0$, although we use its principal value formulation here.
That is, using $G_0^{(\pm)}$ and making the replacement $\T \to T^{(\pm)}$ will yield an emulator for $T^{(\pm)}$. 
}
\begin{align} \label{eq:LS}
    \T = V + V G_0 \T,
\end{align}
with the free-space Green's function operator $G_0(E_q)$ at the on-shell energy $E_q = q^2/2\mu$ and reduced mass $\mu$.
The energy dependence is implicit in what follows.
We stress that using the $\T$~matrix is just a convenient choice. In fact, $T^{(\pm)}$ can be emulated by imposing the associated boundary conditions on $G_0$.
Although the LS equation~\eqref{eq:LS} has the formal solution
\begin{equation} \label{eq:LS_formal_sol}
    \T = \left(\mathds{1} - V G_0 \right)^{-1} V,
\end{equation}
evaluating Eq.~\eqref{eq:LS_formal_sol} in a given basis can be prohibitively slow for large-scale Monte Carlo sampling because of the fine (quadrature) grids typically necessary to obtain high-accuracy results. 

Instead of solving the LS equation~\eqref{eq:LS} directly for each sampling vector $\lecs$, we propose a variational approach
starting with a trial $\T$~matrix motivated by EC:
\begin{equation} \label{eq:TritzTrial}
    \Tritz(\coeffs) = \sum_{i=1}^{\NEC}\coeff_i \T_i.
\end{equation}
Here, $\{\T_i \equiv \T(\lecs_i) \}_{i=1}^{\NEC}$ are the exact solutions of the LS equation~\eqref{eq:LS} for the training set $\{\lecs_i\}_{i=1}^{\NEC}$, while $\{\coeff_i\}_{i=1}^{\NEC}$ are a~priori unknown coefficients.\footnote{
The coefficients are not normalized, \ie,  $\sum_{i=1}^{\NEC}\coeff_{i} \neq 1$, as opposed to the Kohn variational approach in Ref.~\cite{Furnstahl:2020abp}.
}
To determine these coefficients at each $\lecs$, we apply Newton's variational method~\cite{newton2002scattering,Rabitz:1973nm}, which provides a stationary approximation to the exact scattering $\T$~matrix using the functional
\begin{equation} \label{eq:LS_identity}
\begin{split}
    \Tfunc[\Tritz] &= V + V G_0 \Tritz + \Tritz G_0 V \\ & \quad - \Tritz G_0 \Tritz + \Tritz G_0 V G_0 \Tritz,
\end{split}
\end{equation}
given a trial matrix $\Tritz$ such as the one in Eq.~\eqref{eq:TritzTrial}.
The functional~\eqref{eq:LS_identity} is stationary about exact solutions of the LS equation, \ie, $\Tfunc[\T+\delta\T] = \T + (\delta\T)^2$.

In practice, we determine the stationary solution of the functional~\eqref{eq:LS_identity} in a chosen basis and emulate the scattering $\T$~matrix as the matrix element $\!\mel{\phi'}{\T}{\phi}$. For example, one could choose $\ket{\phi}$ to be a plane-wave partial-wave basis $\ket{k\ell m}$ with momentum $k$ and angular momentum quanta $(l,m)$, or one could keep the angular dependence explicit via $\ket*{\phi}=\ket*{\mathbf{k}}$ in a single-particle basis.
We are interested in emulating $\T$ at the on-shell energy $E_q$, so then $k = k' = q$ for $\ket{\phi}$ and $\bra{\phi'}$.
Expressed in the chosen basis, simplifying the functional~\eqref{eq:LS_identity} after inserting~\eqref{eq:TritzTrial} yields
\begin{align}
    \!\mel{\phi'}{\Tfunc(\lecs, \coeffs)}{\phi} = {\!\mel{\phi'}{V(\lecs)}{\phi}} + \coeffs^\trans \vec{m}(\lecs) - \frac{1}{2} \coeffs^\trans  M(\lecs) \coeffs, \label{eq:LS_identity_beta}
\end{align}
with
\begin{align}
    m_i(\lecs) & = \bra{\phi'} \left[\T_i G_0 V(\lecs) + V(\lecs) G_0 \T_i\right] \ket{\phi}, \label{eq:m_vec} \\
    M_{ij}(\lecs) & = \bra{\phi'} [
                       \T_i G_0 \T_j - \T_i G_0 V(\lecs) G_0 \T_j \notag \\
    & \hspace{0.29in} + \T_j G_0 \T_i - \T_j G_0 V(\lecs) G_0 \T_i
    ] \ket{\phi}. \label{eq:M_mat}
\end{align}
If the potential $V(\lecs)$ is linear in the parameter vector $\lecs$,
then $\vec{m}$ and $M$
can be efficiently reconstructed by linear combinations of matrices pre-computed during the training phase of the emulator.
This results in substantial improvements in CPU time, \eg, for chiral nucleon-nucleon (NN) interactions.

By imposing the stationary condition $\dd \Tfunc /\dd \coeffs = 0$, one then finds
$\coeffsopt(\lecs)$ such that $M \coeffsopt = \vec{m}$.
Given that the optimal $\coeffsopt(\lecs)$ yields a trial matrix~\eqref{eq:TritzTrial} with an error $\delta\T$, we insert $\coeffsopt$ in Eq.~\eqref{eq:LS_identity_beta} to obtain an error $(\delta\T)^2$.
The resulting emulator $\Tfuncopt(\lecs) \equiv \Tfunc(\lecs, \coeffsopt)$ is then
\begin{align} \label{eq:emulator}
    \mel{\phi'}{\T}{\phi} &\approx
    {\!\mel{\phi'}{\Tfuncopt}{\phi}} =
    {\!\mel{\phi'}{V}{\phi}} + \frac{1}{2} \vec{m}^\trans M^{-1} \vec{m}.
\end{align}
Equations~\eqref{eq:m_vec}--\eqref{eq:emulator} are the main expressions for emulating scattering observables
with short-range interactions.
We extend the implementation to the long-range Coulomb potential in Sec.~\ref{sec:coulomb}.%
\footnote{%
Emulating the wave function could follow from working out $\ket{\psi(\lecs)} = \ket{\phi} + G_0 \Tfuncopt(\lecs)\ket{\phi}$, where again the appropriate boundary conditions are implied by the choice of $G_0$~\cite{newton2002scattering,taylor2006scattering}.
}

The EC-motivated trial $\T$~matrix~\eqref{eq:TritzTrial} causes increasingly ill-conditioned matrices $M$ with increasing number of training points $n_t$. To control the numerical noise in the evaluation of Eq.~\eqref{eq:emulator}, we follow Ref.~\cite{Furnstahl:2020abp} and add the regularization parameter $\eta = 10^{-12}$ to the diagonal elements $M_{ii}$. This is a relatively simple yet effective approach compared to other regularization methods~\cite{engl1996regularization}. 

Besides numerical instabilities, Newton's variational method can also exhibit spurious singularities~\cite{Apagyi1991}, similar to the so-called Kohn (or Schwartz) anomalies~\cite{PhysRev.124.1468,PhysRevA.40.6879} observed in applications of the Kohn variational principle~\cite{Kohn:1948col, nesbet1980variational}. For instance, we expect spurious singularities to occur at energies where $M$ is singular, \ie, when there is no (unique) stationary approximation to the $\T$~matrix due to the functional~\eqref{eq:LS_identity}. Different methods to mitigate these singularities have been proposed in the literature~\cite{Ladanyi1988, Winstead1990}. Recently, Ref.~\cite{Drischler:2021qoy} demonstrated that an EC-driven emulator that assesses the consistency of results obtained from a set of Kohn variational principles (with different boundary conditions) is effective in detecting Kohn anomalies. If detected, they can be mitigated at a given energy, \eg, by changing the number of training points used for emulation. A similar approach could be applied here; however, we have not encountered issues related to spurious singularities in our comprehensive proof-of-principle calculations presented in Sec.~\ref{sec:results}.

In \verifyvalue{Appendix~\ref{sec:gradients}}, we show that our approach also allows for gradients with respect to model parameters to be straightforwardly propagated. This is an important feature since many optimization and sampling algorithms require gradients.
In \verifyvalue{Appendix~\ref{sec:pw_GreensF}}, we discuss a simple and computationally efficient method to evaluate products involving Green's functions in the partial-wave basis.

\section{Results}\label{sec:results}

Throughout this section, we use the convention that $K_\ell = - \tan\delta_\ell$ (which is opposite to Ref.~\cite{Furnstahl:2020abp}) with all factors of $\pi$, the reduced mass, and momentum accounted for.

\subsection{The Minnesota Potential} \label{sec:minnesota}

\begin{figure}[tb]
    \centering
    \includegraphics{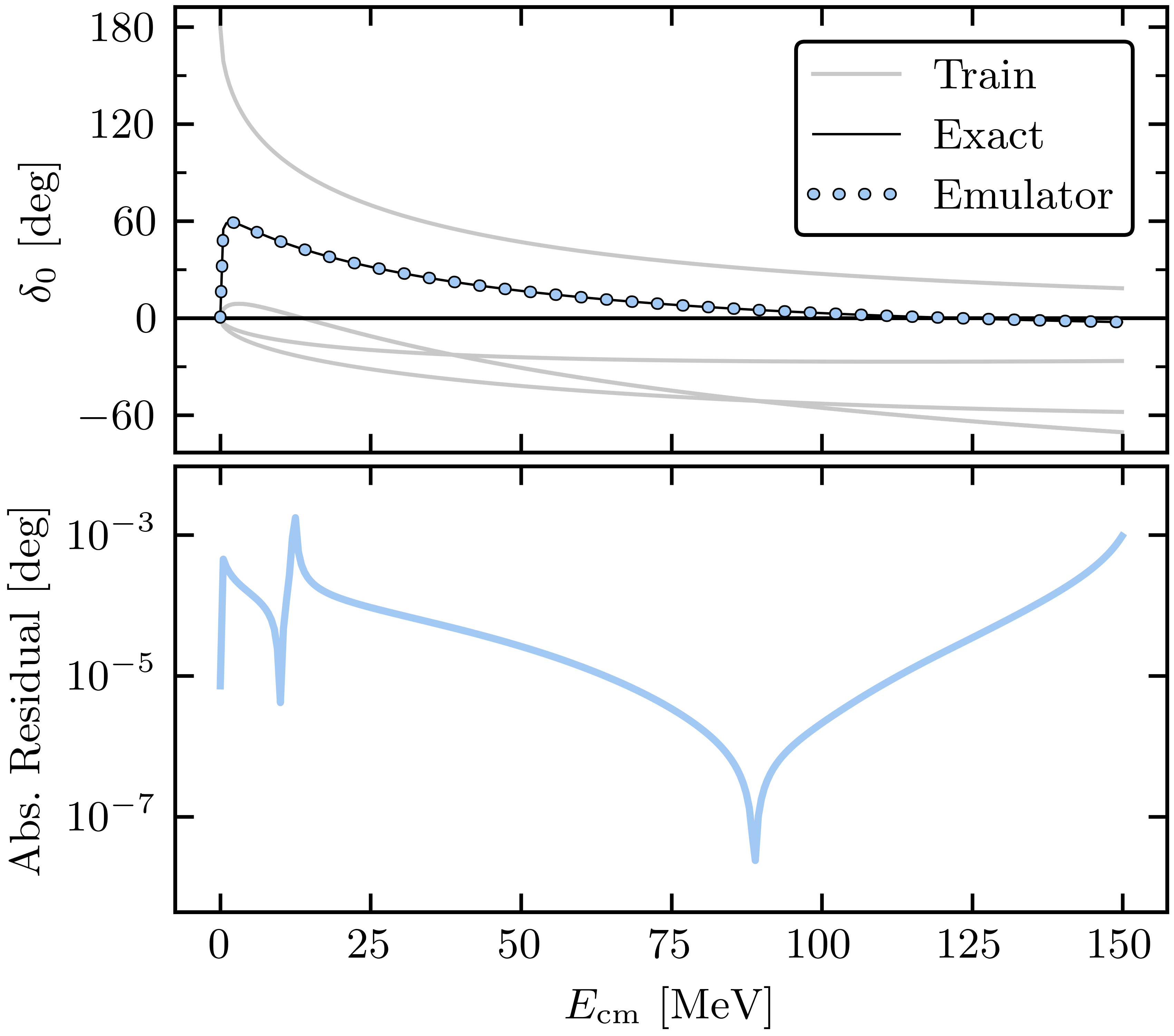}
    \caption{%
    Nucleon-nucleon ${^1}S_0$ phase shifts $\delta_0$ (top panel) and their absolute residuals relative to the exact solutions (bottom panel) for the Minnesota potential~\cite{THOMPSON197753}.
    The training parameters consist of four sets of potential depths in Eq.~\eqref{eq:minnesota}, and the exact phase shifts corresponding to these training points are depicted by the gray lines.
    For details, see the main text.
    }
    \label{fig:minnesota_1S0_phases}
\end{figure}

Following Ref.~\cite{Furnstahl:2020abp}, we apply our emulator to the Minnesota potential~\cite{THOMPSON197753} in the ${^1}S_0$ channel as a simple test case:
\begin{equation} \label{eq:minnesota}
    V(r) = V_{0R} e^{-\kappa_R r^2} + V_{0s} e^{-\kappa_s r^2}.
\end{equation}
The values best reproducing NN scattering phase shifts are $V_{0R} = 200\MeV$ and $V_{0s} = -91.85 \MeV$, as well as $\kappa_R = 1.487\fmis$ and $\kappa_s = 0.465\fmis$~\cite{THOMPSON197753}.
We use the same parameter set as in Ref.~\cite{Furnstahl:2020abp} for training, \ie, $(V_{0R}, V_{0s}) = \{ (0., -291.85)$, $(100., 8.15)$, $(300., -191.85)$, $(300., 8.15) \}$ in units of $\MeV$, and keep $\kappa_R$ and $\kappa_s$ fixed at their best values. Figure~\ref{fig:minnesota_1S0_phases} shows the emulated ${^1}S_0$ phase shifts (top panel) and the absolute residuals (bottom panel) as a function of the center-of-mass energy
at the best-fit values.
For comparison, the phase shifts corresponding to exact solutions of the LS equation~\eqref{eq:LS} for the 4 training points are depicted as gray lines.
The emulated phase shifts reproduce well the exact results, as quantified by the absolute residuals in the bottom panel.

\begin{figure}[tb]
    \centering
    \includegraphics{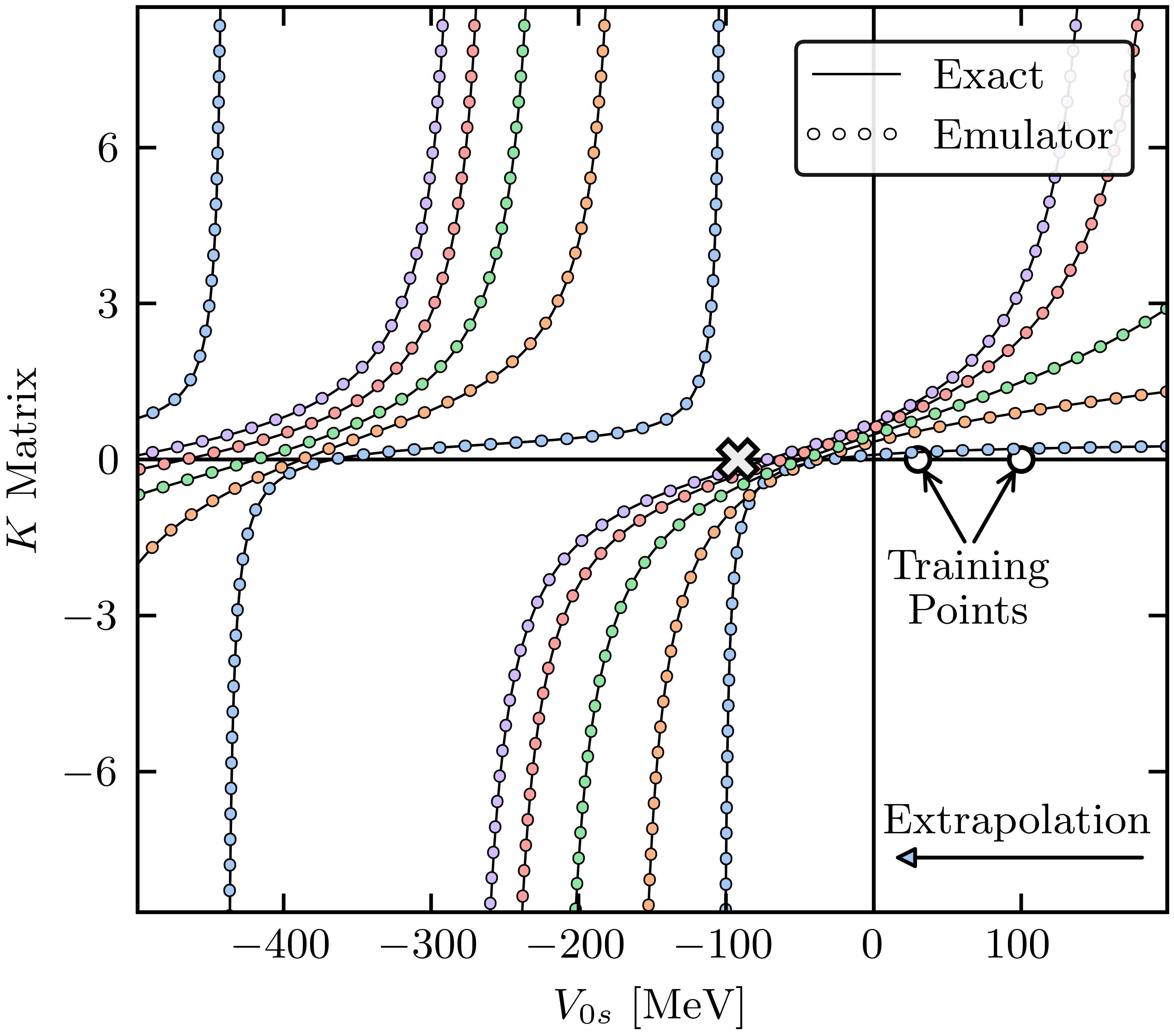}
    \caption{%
    The emulator as a robust tool for extrapolation: on-shell $\T$~matrix in the ${^1}S_0$ channel at fixed center-of-mass energies as a function of the Minnesota potential parameter $V_{0s}$.
    The exact solutions of the LS equation are shown as black lines and the emulator predictions as colored dots.
    Each set of colored dots corresponds to a specific center-of-mass energy (blue: $\verifyvalue{1} \MeV$, orange: $\verifyvalue{15} \MeV$, green: $\verifyvalue{30} \MeV$, red: $\verifyvalue{50} \MeV$, and purple: $\verifyvalue{70} \MeV$).
    Only the two training points, depicted by the open circles, were used. The cross marks the location of the best value for $V_{0s}$ corresponding to Fig.~\ref{fig:minnesota_1S0_phases}. 
    See the main text for more details.
    }
    \label{fig:minnesota_1S0_extrapolation_K}
\end{figure}

We now demonstrate that the emulator is also a robust tool for extrapolations.
We set the Minnesota potential parameters $(V_{0R}, \kappa_R,\kappa_s)$ to their best fit values, train on only two purely repulsive parameter sets with $V_{0s} = 30$ and $100\MeV$, and then extrapolate to purely attractive potentials (capable of supporting bound states).
Figure~\ref{fig:minnesota_1S0_extrapolation_K} shows the resulting on-shell $\T$ matrices in the ${^1}S_0$ channel obtained using our emulator (colored dots) in comparison to the exact solutions of the LS equation (black lines).
Each set of colored dots corresponds to a specific center-of-mass energy in the range $1$--$70 \MeV$ (see the legend for details).
The two training points are depicted by the open circles, and the cross marks the location of the best-fit value for $V_{0s}$.
As the figure illustrates, the emulator can accurately extrapolate far away from the two training points even after passing through poles in both $\T$ and $\T^{-1}$.

\subsection{Including the Coulomb Interaction}
\label{sec:coulomb}

Long-range interactions, such as the Coulomb interaction, are problematic for the LS equation approach whether or not an emulator is employed.
Nevertheless, we can include the Coulomb interaction 
via the Vincent-Phatak method~\cite{Vincent:1974zz,Lu:1994yw}.
The basic idea is to cut off the Coulomb potential at a finite radius so that Eq.~\eqref{eq:LS} applies and then restore this physics using a matching procedure.
Specifically, we emulate the $\T$ matrix from the potential $V^{r_c}(r, r') = V_s(r, r') + V_C^{r_c}(r) \delta(r-r')/(r r')$, where $V_s(r, r')$ is a (non-local) short-range potential and
\begin{align}
    V_C^{r_c}(r) = V_C(r) \theta(r_c - r) 
\end{align}
is the Coulomb potential cut off at a radius $r_c$
large enough such that the short-range potential $V_s$ is negligible.
The modified potential $V^{r_c}(r, r')$ is short-ranged and hence compatible with Eqs.~\eqref{eq:m_vec}--\eqref{eq:emulator}; this is the potential we use to train the emulator.

Suppose we want to emulate the $\T$~matrix in an uncoupled partial-wave channel with angular momentum $\ell$.
Solving Eq.~\eqref{eq:emulator} with $V^{r_c}_\ell$ yields the associated $\T_\ell^{r_c}$, but this is an artificial quantity representing the phase shift relative to the free radial wave functions, \ie,
\begin{equation} \label{eq:asymLimFree}
    u_\ell^{r_c}(q, r) = \regfreewf_\ell(qr) + \T_\ell^{r_c} \irrfreewf_\ell(qr), \qq{for} r \geqslant r_c,
\end{equation}
expressed in terms of the regular $\regfreewf_\ell(qr)$ and irregular $\irrfreewf_\ell(qr)$ Riccati-Bessel function.
To obtain the phase shifts with respect to the Coulomb wave functions 
(Sommerfeld-parameter dependencies being implicit), \ie, 
\begin{equation}\label{eq:asymLimCoul}
    u_\ell(q,r) = \regwf_\ell(qr) + \T^C_\ell \irrwf_\ell(qr), 
    \qq{for} R \leqslant r \leqslant r_c,
\end{equation}
we match the logarithmic derivatives of Eqs.~\eqref{eq:asymLimFree} and~\eqref{eq:asymLimCoul} at $r=r_c$. 
(Here $R$ is the range of the short-range potential.)
This amounts to computing
\begin{align}
    \T_\ell^C = -\frac{\regwf_\ell - A_\ell \regwf_\ell'}{\irrwf_\ell - A_\ell \irrwf_\ell'}, \qq{where}
    A_\ell = \frac{\regfreewf_\ell + \irrfreewf_\ell \T_\ell^{r_c}}{\regfreewf_\ell' + \irrfreewf_\ell' \T_\ell^{r_c}}
\end{align}
and primes denote derivatives with respect to $r$.
Now, both $\T_\ell^C = -\tan\delta_\ell^C$ and the phase shift $\delta_\ell^C$ are with respect to the Coulomb wave functions such that the dependence on the choice of $r_c$ has been removed.
Note that the above relies on a sign convention where, \eg, $\irrfreewf_0 \sim  - \cos(qr)$ and similarly for $\irrwf_0$.
Each of $\regfreewf_\ell$, $\irrfreewf_\ell$, $\regwf_\ell$, $\irrwf_\ell$ and their derivatives can be computed once and stored for emulation purposes.
Solving for $\T^C_\ell$ need only be performed for the on-shell matrix and is a quick post-processing step for the emulator.

\begin{figure}[tb]
    \centering
    \includegraphics{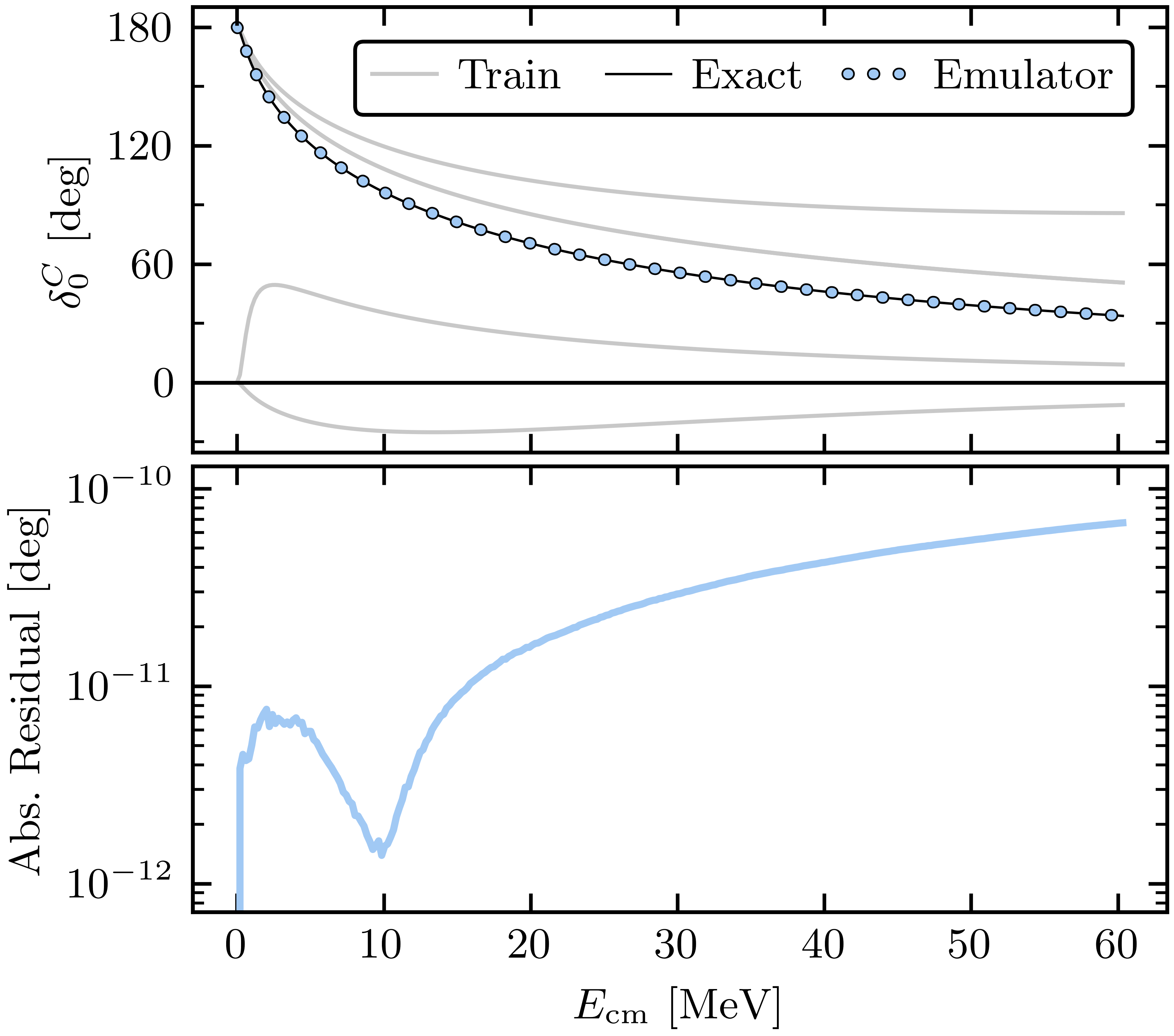}
    \caption{%
    Proton-alpha ${^1}S_0$ phase shifts $\delta_0^C$ with respect to Coulomb wave functions for the non-local potential~\eqref{eq:palphaPotential} (top panel).
    Notice the small absolute residuals (bottom panel) that were obtained with only four training points. For more details, see the caption of Fig.~\ref{fig:minnesota_1S0_phases} and the main text.
    }
    \label{fig:coulomb_p_alpha_phases_and_residuals}
\end{figure}

We apply this approach, with $r_c = 20$\,fm, to proton-$\alpha$ scattering with the non-local potential~\cite{Ali:1984ds}\footnote{
This non-local potential includes a factor of $rr'$ 
consistent with the convention $V(r, r') = V(r)\delta(r-r')$ for a local potential.
Note that $V_{p\alpha,\ell}^{(0)}$ includes a factor of $2\mu$.
}
\begin{equation} \label{eq:palphaPotential}
    V_\ell(r, r') = V_{p\alpha,\ell}^{(0)} r'^{\ell} r^{\ell} e^{-\beta_\ell (r + r')}
\end{equation}
in the $S$-wave; \ie, $\ell=0$.
With the four training points $V_{p\alpha,\ell}^{(0)} = \{\verifyvalue{-30, -1, 1, 10} \}$\,fm$^{-3}$
and $\beta_0 = 0.8$\,fm$^{-1}$, the emulator accurately predicts the phase shift at the optimal value of $V_{p\alpha,\ell}^{(0)} = \verifyvalue{-6.5}$\,fm$^{-3}$~\cite{Ali:1984ds}, as shown in Fig.~\ref{fig:coulomb_p_alpha_phases_and_residuals}.
Across the energy range shown in the figure, $E_\text{cm} \leqslant 60 \MeV$,
the absolute residuals in the phase shift emulator (bottom panel) are negligible. The high accuracy obtained is remarkable because 
computing the phase shift at each energy only involves inverting a $4\times 4$ matrix.

\subsection{Coupled Channels}
\label{sec:coupled_channels}

\begin{figure*}[tb]
    \centering
    \includegraphics{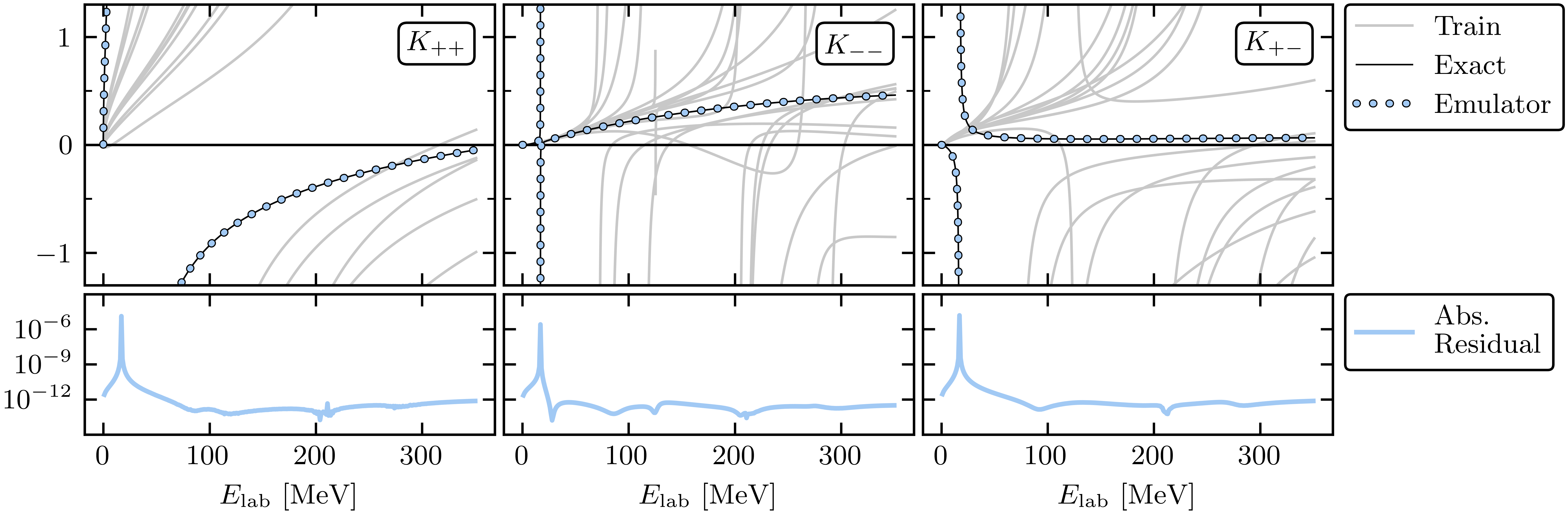}
    \caption{%
    The partial-wave components of the on-shell $\T$ matrix for the $^3S_1$--$^3D_1$ coupled channel in $np$ scattering.
    From left to right: pure $D$-wave, pure $S$-wave, and mixed $S$--$D$-wave component.
    The predictions are made using the semilocal chiral NN potential \verifyvalue{\NNNNLOp SMS} regularized with momentum cutoff $\Lambda = 450\MeV$, which depends on \verifyvalue{6} parameters.
    We use \verifyvalue{12} randomly selected $\lecs$ values in $\verifyvalue{[-5, 5]}$ (in the appropriate units) to train the emulator, whose exact predictions are shown in gray. For more details, see the caption of Fig.~\ref{fig:minnesota_1S0_phases} and the main text.
    }
    \label{fig:np_3S1_3D1_coupled_K_matrix}
\end{figure*}

The straightforward extension to scattering in coupled channels is one of the strengths of our emulator approach. In fact, Eqs.~\eqref{eq:m_vec}--\eqref{eq:emulator} handle them as a special case. Although the term \emph{coupled channels} can also refer to different reaction channels, in the following we specifically consider coupled (spin-triplet) partial-wave channels.

For potentials that are coupled across $n_c$ different channels (\eg, $n_c=2$ for the deuteron), solving the LS equation exactly for one on-shell point requires solving a linear system of dimension $n_c n_k \times n_c n_k$, with $n_k$ being the size of the mesh for an uncoupled channel.
The coupled-channel emulator with $\NEC$ training points instead only involves operations on an $\NEC \times \NEC$ matrix for each desired matrix element of $\T$, where $\NEC \ll n_k$.
Generally, this requires running at most $n_c (n_c+1)/2$ of such emulations because the remaining matrix elements can be determined by symmetry.

We apply this approach to $np$ scattering in the coupled ${^3}S_1$--${^3}D_1$ channel.
The potential used here is the semilocal momentum-space (SMS) regularized chiral potential at \verifyvalue{\NNNNLOp} constructed by Reinert, Krebs, and Epelbaum with momentum cutoff $\Lambda = 450\MeV$~\cite{Reinert:2017usi}.
At this chiral order the ${^3}S_1$--${^3}D_1$ channel depends on \verifyvalue{$\nlecs = 6$} non-redundant parameters, or low-energy constants (LECs), in the NN sector~\cite{Reinert:2017usi}.
We choose $\verifyvalue{\NEC = 2\nlecs = 12}$ training points randomly in the range $[-5, 5]$, where the unit of each parameter is as given in Ref.~\cite{Reinert:2017usi} and left implicit here.
The emulator's predictions are then validated at the best values of the parameters found in Ref.~\cite{Reinert:2017usi}.
We use a compound Gauss--Legendre quadrature mesh of \verifyvalue{80} momentum points to exactly solve the LS equation at the training points.

Figure~\ref{fig:np_3S1_3D1_coupled_K_matrix} shows the resulting on-shell $\T$ matrix obtained from the exact calculation and emulator as a function of the laboratory energy.
Each column corresponds to a different partial-wave component of the $\T$~matrix.
The emulator accurately reproduces the exact $\T$ matrix elements across the wide range of energies shown, $1$--$350\MeV$\@.
Except for a spike near the energy region where the $\T$ matrix is singular, the residuals are on the order of $\verifyvalue{10^{-12}}$. These errors are far beneath the experimental uncertainties if the $\T$ matrix were to be converted to phase shifts~\cite{Perez:2013mwa}.

\subsection{The Scattering Cross Section}

We now combine multiple partial-wave emulators into an overall emulator for nuclear observables.
As a simple example, we show $np$ total cross sections using partial waves up to $j_{\text{max}} = 20$---again with the $\Lambda = 450\MeV$ \verifyvalue{\NNNNLOp SMS} potential~\cite{Reinert:2017usi}, which reproduces well the total cross sections from the partial-wave analysis~\cite{Perez:2013mwa} over a wide range of laboratory energies.
This requires training partial-wave emulators across singlet and triplet channels up to $\verifyvalue{j = 4}$, while the remaining waves are fixed with respect to $\lecs$.
There are a total of 26 free parameters in $\lecs$.
The $\NEC$ training locations are again chosen randomly in $[-5, 5]$, where $\NEC$ is determined based on the $\nlecs$ NN LECs in each partial wave via $\NEC = \max(2\nlecs, 4)$.

\begin{figure}[tb]
    \centering
    \includegraphics{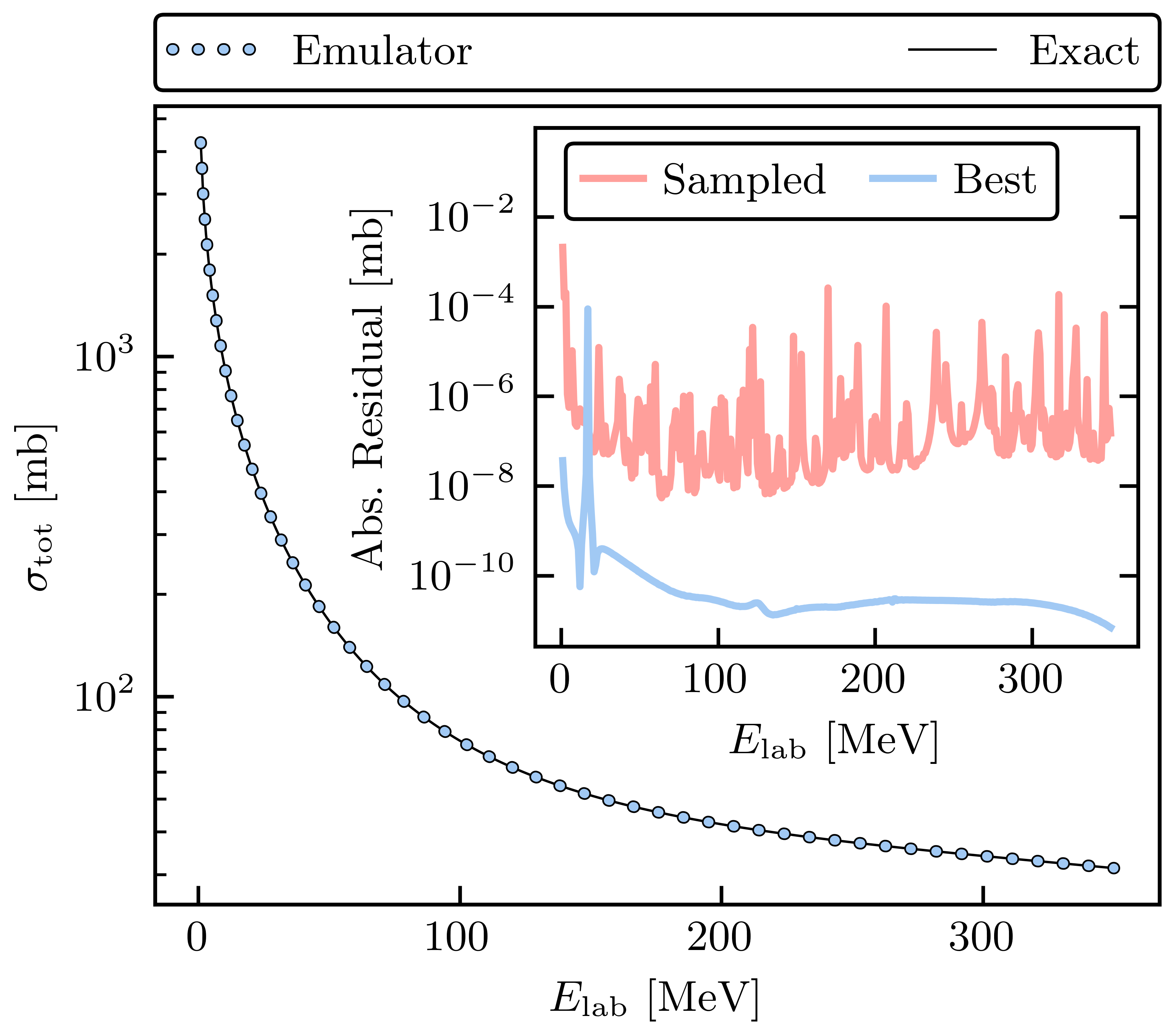}
    \caption{%
    The $np$ total cross section using the $\Lambda = 450\MeV$ \verifyvalue{\NNNNLOp SMS} potential~\cite{Reinert:2017usi} up to $j_{\mathrm{max}}=20$.
    The inset shows the mean absolute error between the emulator and the exact solution for 500 different samples of the NN LECs $\lecs$ (red line) as well as the absolute residual at the best-fit $\lecs$.
    These errors are negligible compared to any experimental uncertainties.
    See the main text for more details.
    }
    \label{fig:np_total_cross_section}
\end{figure}

Upon emulating $\T_j$, the total cross section can be calculated via 
\begin{align}
    \sigma_{\text{tot}}(q) = - \frac{\pi}{2q^2} \sum_{j=0}^{j_{\text{max}}} (2j+1) \Re{ \Tr [S_j(q)-\mathds{1}] },
\end{align}
where $S_j = \mathds{1} + 2i (\mathds{1} + i \T_j)^{-1} \T_j$ and $q$ is the center-of-mass momentum.
Both $S_j$ and $\T_j$ are $4\times 4$ matrices that contain both the triplet-triplet and the singlet-triplet channels.
Figure~\ref{fig:np_total_cross_section} shows the emulated $\sigma_{\text{tot}}$ at the optimal values of $\lecs$ determined in Ref.~\cite{Reinert:2017usi}.
The emulator has an error $\lesssim \verifyvalue{10^{-10}}$\,mb for $E_{\mathrm{lab}} > 50$\,MeV\@.
For $E_\mathrm{lab} < 50$\,MeV, the error is $\lesssim \verifyvalue{10^{-8}}$\,mb except for the spike due to the singular $\T$~matrix in the $^3S_1$--$^3D_1$ channel, as discussed in Sec.~\ref{sec:coupled_channels}. 
In either case, these errors are vanishingly small compared to both the size of the cross section itself and its experimental uncertainty~\cite{Perez:2013mwa}.

When randomly sampling \verifyvalue{500} values of the NN LECs in the range of $[-15, 15]$---an extrapolation of $\pm 10$
beyond the range of the training data (in the appropriate units)---the average absolute emulator error is less than \verifyvalue{$10^{-7}$\,mb}.
Furthermore, the emulator provides a factor of \verifyvalue{$>300$x} improvement in terms of CPU time relative to the exact calculation.
If the size of the momentum mesh used in the LS equation is increased from 80 to 160 quadrature points, then the factor becomes \verifyvalue{$>1000$x}.
Further acceleration can be expected as finer momentum meshes are used in solving the LS equation, and as $\lecs$ enter into higher partial waves at higher chiral orders.

Meaningful comparisons of the speed and accuracy obtained here with the emulators in Refs.~\cite{Furnstahl:2020abp,Drischler:2021qoy} requires, at least, that all calculations be performed in the same space. 
For this benchmark, we have implemented the Kohn variational principle with uncoupled channels in momentum space. Our findings so far indicate that the two variational methods are comparable in accuracy (for the same quadrature rule) for the chiral potential, although the relative speedups are implementation dependent. 
More work is necessary to provide more quantitative comparisons. 

We also compare the accuracy of our emulator to a promising accelerator for NN scattering observables developed in Ref.~\cite{Miller:2021pcu}.
Instead of a variational method, Miller~\etal\ employed the wave-packet continuum discretization (WPCD) method to approximate scattering solutions at multiple energies at once. This method is well-suited for parallelization using Graphics Processing Units, which can lead to significant speedups compared to exact calculations via conventional matrix inversion. Depending on the laboratory energy and the number of wave packets included, Miller~\etal\ reported averaged errors in the total cross section on the order of $1$\,mb at best based on the chiral interaction NNLO$_\text{opt}$~\cite{Ekstrom:2013kea}. These errors suggest that our emulator motivated by EC can provide significantly higher accuracies, even when only a few training points per partial-wave channel are used. A quantitative comparison of the methods' efficiencies, however, would require a scattering scenario with matching nuclear interactions.

\section{Summary and outlook}
\label{sec:summary}

We showed that Newton's variational method combined with ideas from eigenvector continuation allows for the construction of a fast \& accurate emulator for two-body scattering observables.
Our approach begins with a trial $\T$ or $T$ matrix constructed from a small number of exact solutions to the LS equation in the parameter space of the Hamiltonian.
Subsequent emulation only requires linear algebra operations on low-dimensional matrices. 

We then provided several applications to short-range potentials with and without the Coulomb interaction and partial-wave coupling.
In all cases studied, the emulator is capable of reproducing phase shifts and total cross sections with remarkable accuracy, even far from the support of the training data and across poles in $\T$ and $\T^{-1}$.
In particular, for a modern chiral interaction at $\NNNNLOp$ the emulator reproduced the exact neutron-proton cross section with negligible error but was over 300x faster in CPU time.
The code that generates all results and figures within this Letter
\verifyvalue{will be made} publicly available~\cite{BUQEYEgithub}.

While the number of emulators applicable to bound-state observables in few- and many-body systems is growing~\cite{Konig:2019adq,Demol:2019yjt,Ekstrom:2019lss,Wesolowski:2021cni}, developing methods with similar efficacy for three- and higher-body scattering is an important avenue. 
Thanks to emulators, simultaneous Bayesian fits of chiral interactions to pion-nucleon, nucleon-nucleon, and three-nucleon observables~\cite{Wesolowski:2021cni} with theoretical uncertainties rigorously quantified~\cite{Wesolowski:2018lzj,Melendez:2019izc} already have become feasible.
Next-generation emulators have the potential to extend these studies to three- and higher-body scattering observables and to shed light on important issues inherent in chiral EFT\@.
Our approach, together with the advances made in applying Kohn variational principles based on EC trial wave functions to three-body scattering~\cite{Zhang:2021xx}, is promising in this direction.
Further, the fast convergence we observed with EC-inspired trial matrices (instead of wave functions) motivates the exploration of the EC concept applied to stationary functionals in a more general context.
Altogether, these are exciting prospects for rigorous Bayesian uncertainty quantification in nuclear physics and reaction theory.


\begin{acknowledgments}
We thank Evgeny Epelbaum for sharing a code that generates the SMS chiral potentials and Kyle Wendt for fruitful discussions. We are also grateful to the organizers of the (virtual) INT program ``Nuclear Forces for Precision Nuclear Physics'' (INT--21--1b) for creating a stimulating environment to discuss eigenvector continuation and variational principles.
This work 
was supported in part by the National Science Foundation under Grant No.~PHY--1913069 and the NSF CSSI program under award
number OAC-2004601 (BAND Collaboration~\cite{BAND_Framework}), and the NUCLEI SciDAC Collaboration under U.S. Department of Energy MSU subcontract RC107839-OSU\@.
This material is based upon work supported by the U.S. Department of Energy, Office of Science, Office of Nuclear Physics, under the FRIB Theory Alliance award DE-SC0013617. 
\end{acknowledgments}


\appendix

\section{Gradients}
\label{sec:gradients}

Gradients of predictions with respect to the input parameters are useful for various optimization and Monte Carlo sampling algorithms.
For this reason, and because their form is quite simple, we provide the emulator gradients here.
Consider one parameter $\lec_k$ of the potential $V(\lecs)$.
Then, from Eqs.~\eqref{eq:m_vec}--\eqref{eq:emulator} we have
\begin{align} \label{eq:emulator_gradient}
    \pdv{\lec_k}\mel{\phi'}{\Tfuncopt}{\phi} & =
    \pdv{\lec_k}\mel{\phi'}{V}{\phi} \notag \\
    & + \frac{1}{2} \pdv{\vec{m}^\trans}{\lec_k} M^{-1} \vec{m}
    + \frac{1}{2} \vec{m}^\trans M^{-1} \pdv{\vec{m}}{\lec_k}
     \\
    & - \frac{1}{2} \vec{m}^\trans M^{-1} \pdv{M}{\lec_k} M^{-1} \vec{m}, \notag
\end{align}
where
\begin{align}
    \pdv{m_i}{\lec_k} & = \bra{\phi'} [\T_i G_0 \pdv{V}{\lec_k} + \pdv{V}{\lec_k} G_0 \T_i] \ket{\phi} \\
    \pdv{M_{ij}}{\lec_k} & = - \bra{\phi'} [\T_i G_0 \pdv{V}{\lec_k} G_0 \T_j + \T_j G_0 \pdv{V}{\lec_k} G_0 \T_i] \ket{\phi}.
\end{align}
The value of $\coeffsopt = M^{-1}\vec{m}$ must already be computed for the emulator itself and thus can be reused in Eq.~\eqref{eq:emulator_gradient}.
The $M$ matrix is symmetric, which means that $\vec{m}^\trans M^{-1} = (M^{-1}\vec{m})^\trans$ requires no further computation.

If $V(\lecs)$ is linear in $\lecs$, then each projection of the gradient tensors $\partial \vec{m}/\partial \lec_k$ and $\partial M/\partial \lec_k$ can be performed once and stored.
Therefore, all components of the gradient at $\lecs$ can either be pre-computed during training or have already been completed during the emulation step for $\T(\lecs)$---only matrix multiplication remains to be done to obtain $\partial \T/\partial \lec_k$.
Thus, not only is the emulator fast to compute, the gradient of the emulator is also fast because it operates in the small space of training points.
This makes gradients feasible to incorporate into sampling codes with little computational overhead.

\begin{figure}[tb]
    \centering
    \includegraphics{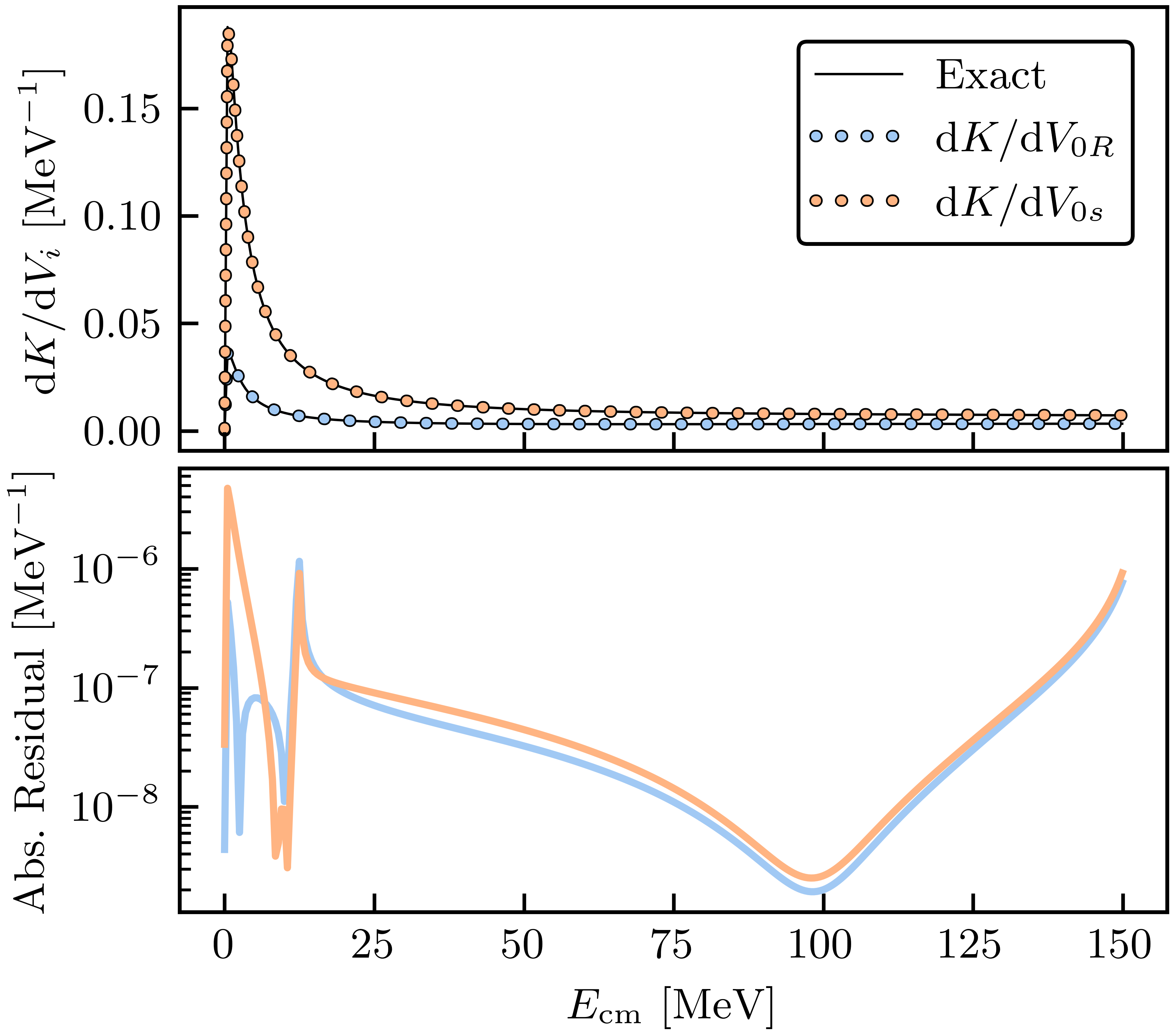}
    \caption{%
    The exact and emulated gradients of the $\T$ matrix with respect to the Minnesota potential parameters $V_{0R}$ (in blue) and $V_{0s}$ (in orange).
    The prediction is made at the best fit values of $V_{0R} = 200\MeV$ and $V_{0s} = -91.85 \MeV$~\cite{THOMPSON197753}.
    The emulator uses the same four training locations as in Sec.~\ref{sec:minnesota} and is able to reproduce the exact gradients to within $10^{-6}\MeV^{-1}$ for almost all center-of-mass energies (bottom panel).
    }
    \label{fig:minnesota_1S0_grads}
\end{figure}

As an example, Fig.~\ref{fig:minnesota_1S0_grads} shows the gradients of the $\T$ matrix from the Minnesota potential considered in Sec.~\ref{sec:minnesota}---with  the same training points.
The gradient is computed at the best values of $V_{0R}$ and $V_{0s}$; the residuals compared to the exact calculation are 
negligible for all
center-of-mass energies shown.

\section{A Convenient Form for the Partial-Wave Green's Function} \label{sec:pw_GreensF}

The free-space Green's function regularly acts between operators when setting up the emulator, and possibly during each emulation if $V(\lecs)$ cannot be projected and stored up front.
Although a product like $V G_0 \T$ appears straightforward to evaluate, there are technicalities involving numerical instabilities and integration measures that are obscured when the LS equation~\eqref{eq:LS} and the emulator equations~\eqref{eq:m_vec} and~\eqref{eq:M_mat} are written in operator form.
Thus, it can be convenient to construct a form of $G_0$ that can be applied  
as a matrix-matrix product, such that all the aforementioned equations can be evaluated straightforwardly.
This approach has the added benefit that it only requires generating the potential $V$ on the fixed grid used for solving the LS equation~\cite{Wendt:2018private,GlockleInterpolation1982}, rather than appending entries for each specific on-shell solution~\cite{Landau:1996}.
This means that generating $V$ and $\T$ for new parameters $\lecs$ becomes more efficient in both runtime and memory.

When solving the LS equation in partial waves, the following projection arises~\cite{Landau:1996}:
\begin{align}
    \mel{p'\ell m}{V G_0 \T}{p \ell m} = \frac{2}{\pi} \mathcal{P}\!\!\int_0^{\infty} \!\!\dd{k} k^2 \frac{V_{\ell}(p', k) \T_{\ell}(k, p; q)}{(q^2 - k^2)/2\mu},
\end{align}
where we work in uncoupled channels for simplicity and $\mu$ is the reduced mass.
The $\mathcal{P}$ denotes the principal value integral due to our choice of $G_0$.
To avoid the numerical instability (\ie, pole) at $k=q$, a zero integral is subtracted to yield
\begin{align}
    & \mel{p'\ell m}{V G_0 \T}{p \ell m} \\
    & = \frac{2}{\pi}\! \int_0^{\infty} \!\!\dd{k}\frac{V_\ell(p', k) \T_\ell(k, p; q) k^2 - V_\ell(p', q) \T_\ell(q, p; q) q^2}{(q^2 - k^2)/2\mu}. \notag
\end{align}
See also Ref.~\cite{Hoppe:2017lok} for a similar numerical approach.
This can be compressed by writing
\begin{align}
    \mel{p'\ell m}{V G_0 \T}{p \ell m} = \frac{2}{\pi} \int V_\ell(p', k) \dd{G_0(k; q)} \T_\ell(k, p; q),
\end{align}
where, in the partial-wave basis,
\begin{align}
    \dd{G_0(k; q)} = \frac{2\mu k^2\dd{k}}{q^2-k^2} - \dd{k}\delta(k-q) \int_0^{\infty} \dd{p} \frac{2\mu q^2}{q^2-p^2}.
\end{align}
The left-hand term is the standard free-space Green's function with a factor of $k^2\dd{k}$ included.
The right-hand term is the zero integral (in principal value) included for numerical stability, which has been multiplied by a factor that will get the on-shell portion of whatever matrices it acts upon.

In practice, the potential---and hence $\T$---must be evaluated on a grid using, \eg, Gauss--Legendre quadrature for $k$ and $\dd{k}$.
This means that the $\dd{k}\delta(k-q)$ will not be able to set the gridded matrix elements to $k=q$.
Thus, we replace $\dd{k}\delta(k-q)$ with an interpolation vector $S(q)$ that performs the mapping
\begin{align}
    \sum_k f(k) S_k(q) \to f(q)
\end{align}
for any smooth $f$ that has been evaluated on some grid of $k$, as in Ref.~\cite{GlockleInterpolation1982}.
Because both the quadrature grid in $k$ and the required on-shell locations $q$ are fixed throughout the emulation process, $S_k(q)$ needs only to be computed once.
Therefore, all the components of $\dd{G_0}$ are independent of $\lecs$ and we can avoid unnecessary calculations while sampling.

The resulting matrix $\dd{G_0}$, which is diagonal in momentum space, is what we use as $G_0$ in all emulators shown here.
It reduces the radial integrals over $k$ and singularity smoothing in products like $V G_0 \T$ to a matrix product of $V \dd{G_0} \T$ for matrices on a fixed mesh for $k$ and $\dd{k}$.
Another benefit of this approach is the simplicity it brings to the exact solutions of the LS equation: we can now use $\dd{G_0}$ in Eq.~\eqref{eq:LS_formal_sol}.
Upon solving for $\T_\ell(p', p; q)$ via Eq.~\eqref{eq:LS_formal_sol}, one can again use the interpolation vector $S(q)$ to compute the on-shell component: $\T_\ell(q) = S(q)^\trans \T_\ell S(q)$.
This sidesteps the requirement of creating unique $(n_k+1)\times(n_k+1)$ matrices for each desired on-shell $\T_{\ell}(q)$, as espoused in Ref.~\cite{Landau:1996} and elsewhere. The advantage is prominent when computing the different partial waves for the total cross section calculation.


\bibliography{bayesian_refs,refs}

\end{document}

%% file: buqeye_macros.tex
\makeatletter
\newcommand\newsubcommand[3]{\newcommand#1{#2\sc@sub{#3}}}
\def\sc@sub#1{\def\sc@thesub{#1}\@ifnextchar_{\sc@mergesubs}{_{\sc@thesub}}}
\def\sc@mergesubs_#1{_{\sc@thesub#1}}

\newcommand\newsupcommand[3]{\newcommand#1{#2\sc@sup{#3}}}
\def\sc@sup#1{\def\sc@thesup{#1}\@ifnextchar^{\sc@mergesups}{^{\sc@thesup}}}
\def\sc@mergesups^#1{^{\sc@thesup#1}}
\makeatother

\DeclareMathAlphabet{\mathbcal}{OMS}{cmsy}{b}{n}

\newcommand{\etal}{\textit{et~al.}\xspace}

\newcommand{\fmis}{\, \text{fm}^{-2}}

\newcommand{\MeV}{\, \text{MeV}}

\newcommand{\NNNNLOp}{\ensuremath{{\rm N}{}^4{\rm LO}+}\xspace}

\newcommand{\nlecs}{n_a}

\newcommand{\ordervec}{\vec}

\newcommand{\inputvec}{\mathbf}

\newsubcommand{\ckvec}{\ordervec{c}}{k}

\newsubcommand{\bkvec}{\ordervec{b}}{k}

\newsubcommand{\ckvecset}{\ordervec{\inputvec{c}}}{k}

\newsubcommand{\ckvecapprox}{\mathbf{c}'}{k}
\newsubcommand{\ckvecapproxset}{\mathbf{C}'}{k}

\newsubcommand{\bkvecapprox}{\mathbf{b}'}{k}
\newsubcommand{\bkvecset}{\mathbf{B}}{k}
\newsubcommand{\bkvecapproxset}{\mathbf{B}'}{k}

\newcommand{\genobs}{y}

\newsubcommand{\genobsvec}{\ordervec{\genobs}}{k}
\newsubcommand{\genobsvecset}{\ordervec{\inputvec{\genobs}}}{k}

\newcommand{\lecs}{\vec{a}}

\newsubcommand{\akvec}{\mathbf{a}}{k}

\newsubcommand{\akvecapprox}{\mathbf{a}'}{k}
\newsubcommand{\akvecset}{\mathbf{A}}{k}
\newsubcommand{\akvecapproxset}{\mathbf{A}'}{k}

{}  

\newcommand{\trans}{\intercal}

\def\diffd{\mathrm{d}}  

\DeclareDocumentCommand\differential{ o g d() }{ 
    \IfNoValueTF{#2}{
        \IfNoValueTF{#3}
            {\diffd\IfNoValueTF{#1}{}{^{#1}}}
            {\mathinner{\diffd\IfNoValueTF{#1}{}{^{#1}}\argopen(#3\argclose)}}
        }
        {\mathinner{\diffd\IfNoValueTF{#1}{}{^{#1}}#2} \IfNoValueTF{#3}{}{(#3)}}
    }
\DeclareDocumentCommand\dd{}{\differential} 

\newcommand{\pathd}{\mathcal{D}}  

\DeclareDocumentCommand\pathdifferential{ o g d() }{ 
    \IfNoValueTF{#2}{
        \IfNoValueTF{#3}
            {\pathd\IfNoValueTF{#1}{}{^{#1}}}
            {\mathinner{\pathd\IfNoValueTF{#1}{}{^{#1}}\argopen(#3\argclose)}}
        }
        {\mathinner{\pathd\IfNoValueTF{#1}{}{^{#1}}#2} \IfNoValueTF{#3}{}{(#3)}}
    }